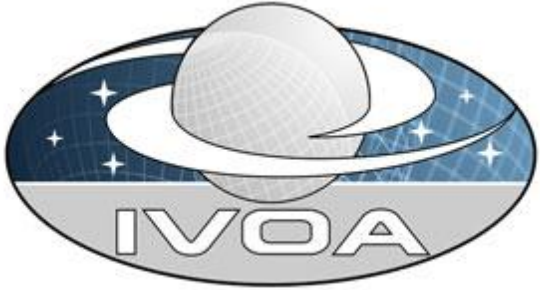

*International*

*Virtual*

*Observatory*

*Alliance*

# IVOA Credential Delegation Protocol

# Version 1.0
*IVOA Recommendation*
*2010 February 18*

**This version:**
   http://www.ivoa.net/Documents/CredentialDelegation/20100218
**Latest version:**
   http://www.ivoa.net/Documents/latest/CredentialDelegation.html
**Previous version(s):**
PR 1.0:   http://www.ivoa.net/Documents/CredentialDelegation/20090818
PR 1.01:  http://www.ivoa.net/Documents/CredentialDelegation/20080923
WD 1.01:  http://www.ivoa.net/Documents/CredentialDelegation/20080915
WD 1.0:   http://www.ivoa.net/Documents/CredentialDelegation/20080715

**Author(s):**
   Matthew Graham
   Raymond Plante
   Guy Rixon
   Giuliano Taffoni

## Abstract

The credential delegation protocol allows a client program to delegate a user's credentials to a service such that that service may make requests of other services in the name of that user. The protocol defines a REST service that works alongside other IVO services to enable such a delegation in a secure manner. In addition to defining the specifics of the service protocol, this document describes how a delegation service is registered in an IVOA registry along with the services it supports. The specification also explains how one can determine from a service registration that it requires the use of a supporting delegation service.

## Status of This Document

This is an IVOA Recommendation made available for public reference. The first release of this document was 2008 September 23.

*This document has been produced by the IVOA Grid and Web Services Working Group.*

*It has been reviewed by IVOA Members and other interested parties, and has been endorsed by the IVOA Executive Committee as an IVOA Recommendation. It is a stable document and may be used as reference material or cited as a normative reference from another document. IVOA's role in making the Recommendation is to draw attention to the specification and to promote its widespread deployment. This enhances the functionality and interoperability inside the Astronomical Community.*

*A list of* current IVOA Recommendations and other technical documents *can be found at* http://www.ivoa.net/Documents/.

## Acknowledgements

The concept of delegation by impersonation was promoted by the grid computing movement and particularly by the Globus project. The protocol described below is derived from the delegation service in Globus Toolkit 4.

This document has been developed with support from the National Science Foundation's Information Technology Research Program under Cooperative Agreement AST0122449 with the John Hopkins University, from the UK Science and Technology Facilities Council (STFC), and from the European Commission's Sixth Framework Program via the Optical Infrared Coordination Network (OPTICON).

## Conformance related definitions

The words "MUST", "SHALL", "SHOULD", "MAY", "RECOMMENDED", and "OPTIONAL" (in upper or lower case) used in this document are to be interpreted as described in IETF standard, RFC 2119 [RFC 2119].

The **Virtual Observatory (VO)** is a general term for a collection of federated resources that can be used to conduct astronomical research, education, and outreach. The **International Virtual Observatory Alliance (IVOA)** is a global collaboration of separately funded projects to develop standards and infrastructure that enable VO applications. The International Virtual Observatory (IVO) application is an application that takes advantage of IVOA standards and infrastructure to provide some VO service.

# Contents

# 1   Introduction

Some services in the VO have restricted access; some users have access rights. When a client program makes a request to one of these secured services, it is typically doing so in the name of the user running the client, i.e. the client presents to the service credentials authenticating the user's identity.

"Presents to the service credentials" means that the client sends the public credentials (an X.509 certificate) but not the private credentials (the private key matching the public key in the certificate). The client authenticates its right to use the identity in the certificate by proving that it holds the private key. It does this using one of the approved authentication methods [1], either Transport Layer Security (TLS) or digital signature.

Now consider a secured service -- call it the "agent" -- which needs to drive other secured services. This might be a "broker" service accessing restricted archives, or it might be a DAL service storing query results in VOSpace. The agent has the certificate for the user's identity but it does not have the private key, so it cannot authenticate the user's identity.

To proceed to operate on the user's behalf, the agent needs to get a private key. Sending the user's own private key across the network is too dangerous and vulnerable to interception. The alternative, on which this IVOA delegation protocol is based, is to generate a *proxy identity* tied to the user's identity, with a certificate based on a key pair generated by the agent. The agent can then authenticate the proxy identity to other services. Those other services recognize a proxy identity by the annotations in its certificate and accord it the same access rights as the primary identity from which the proxy is derived.

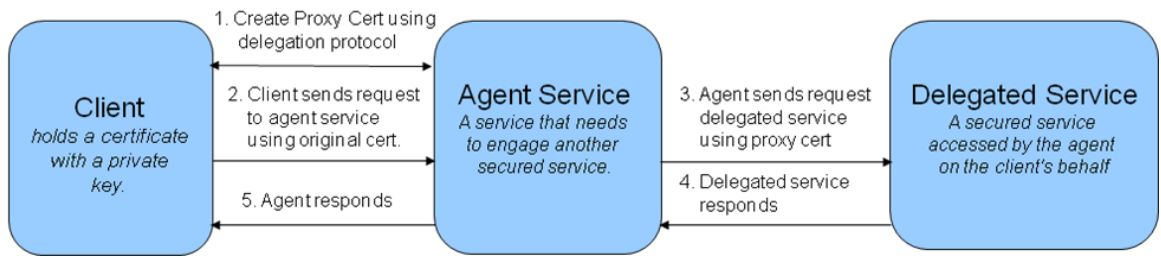

**Figure 1:** Recipe for creating a proxy certificate so that an agent may access another service with the client's identity.

This method of delegation, summarized in Fig. 1, is called "delegation by impersonation" and is common in grid computing [2].

The certificate for a proxy identity is commonly called a proxy certificate or simply "a proxy". IETF RFC 3820 [3] defines the content and encoding of these certificates.

The subject or "distinguished name" (DN) in a proxy certificate is formed by adding an additional CN field to the DN of certificate from which it is derived. The value of the added CN can be any legal DN string, but is typically unique to the usage context (e.g. user session). For example, if an end-entity certificate (EEC; the long-term certificate issued to a user) authenticates the identity represented by the DN,

*C=UK, O=AstroGrid, OU=IoA Cambridge, CN=Guy Rixon*,

then a proxy certificate made from this will authenticate

*C=UK, O=AstroGrid, OU=IoA Cambridge, CN=Guy Rixon, CN=183103.*

To trust the identity in a certificate, an entity must be able to trace a chain of signatures back to a trust anchor. When a proxy certificate is made from an EEC, then the EEC becomes part of this chain. Therefore, whenever a proxy is sent to a service to authenticate an identity, the EEC from which the proxy is derived must also be sent.

A proxy certificate may be derived from another proxy instead of directly from an EEC, in which case it the DN will include multiple added CN fields. This is normal in the VO, since many desktop applications get their credentials as proxies; they never see a private key matching an EEC. Regardless of how many proxies appear in the credentials presented for authentication to the service, the delegation process establishes what we refer to as a *delegated identity* on the server; that delegated identity is represented by a DN that matches the DN in the EEC at the end of the certificate chain.

Note that in IVOA protocols, a client must delegate credentials *before* calling a service that needs to use delegated credentials. The client can find out the need for delegation from the service registration.

The delegation process that provides a delegated identity to the agent has these steps:

1. The client commands the agent to generate and store a key pair for a particular identity.
2. The client retrieves from the agent a certificate signing request (CSR) containing the public key.
3. The client generates from the CSR a certificate, signs it and gives the certificate to the agent.
4. The agent securely stores the certificate.

Once the agent has a proxy certificate, the client is ready to access the secured service. The client authenticates with the service, and the service matches the EEC identity of the client with a stored proxy that it can use on the client's behalf to access other services.

All client requests during the delegation process must be authenticated, and the server must only allow a client to access credentials corresponding to its own delegated identity. In principle, authentication can be accomplished either via the *TLS-with-client-certificate method* [11], i.e. by HTTPS, or via digital signatures on the message bodies (in order to be compliant with the IVOA authentication standard [1]). This version of the delegation protocol standard specifically requires the TLA-based method on HTTPS URLs using RESTful operations. A future version may address a SOAP-based interface using digital signatures whose design will follow very close to the one outlined here.

## 2   Delegation protocol

### 2.1  Required web resources

#### 2.1.1  Specification

The delegation protocol is RESTful and is enabled via a set of four service components that are each accessible via a URL which we refer to as *web resources,* each representing information accessible to the client through the course of the delegation process.  In particular, an agent that needs to receive proxy credentials must provide four types of web resources accessible by HTTP that represent the following:

- WR1.   a single resource per delegation agent representing a list of the available delegated identities.
- WR2.   a delegated identity, one for each in the identities list.
- WR3.   a certificate signing request (CSR), one for each identity.
- WR4.   a proxy certificate, one for each identity.

Only WR1, the list of delegated identities, is represented by a static URL, always available as long as the delegation agent is on-line.  The others are created dynamically through the course of the protocol.  In this document, a web resource is said to *exist* or be *available* if a properly authenticated and authorized GET request to the resource's URL will return a successful response.

The form of the URL for WR1 can have any path component (the portion following the server's domain name [10]) but it must not include a query component (i.e. a component following a question mark, "?") nor a fragment component (I.e. a component following a pound sign, "#").

> **Example:** Legal and illegal URLs for the list of delegated identities (WR1):
> **Legal:**
> http://mysite.net/cgi-bin/delegationService
> http://mysite.net/delegations
> **Illegal:**
> http://mysite.net/cgi-bin/delegationService?res=data
> http://mysite.net/delegations#list

The forms of the URLs for the other resources reflect parent-child relationship between the resources.  The URL of a so-called *child resource* is composed of the URL of the *parent resource* appended with an additional path. The URLs of these resources must satisfy the following constraints on their paths:

- WR2, a delegated identity, must be a child of WR1, the list of identities.
- WR3, the CSR, must be a child of its delegated identity (WR2), and the additional path must be set to *CSR*.
- WR4, the certificate, must be a child of its delegated identity (WR2), and the additional path must be set to *certificate.*

> **Example:**
> A successful delegation process might interact with the following four specific resources:
> 1. delegated identities list: **https://mysite.net/delegations**
> 2. the delegated identity: **https://mysite.net/delegations/A328FE83D0**
> 3. the CSR: **https://mysite.net/delegations/A328FE83D0/CSR**
> 4. the proxy certificate: **https://mysite.net/delegations/A328FE83D0/certificate**

> **Note:**
> The resources for the identities can have any name, but that name should be easy to encode into the URL; X.500 DNs, as taken from the certificates, are hard to read when URL-encoded and evoke privacy concerns if incorporated into URLs (see next section). A better choice is a hash of the DN that is unique within the delegation service. Java implementations that cache an object for each identity can use the result of the *hashcode()* method to name the identity resource.

## *2.2 Representations of resources*

When one of the web resources is accessed via a GET operation on its URL, it returns a representation of the resource. This section describes those representations.

There is no required representation for WR1, the list of delegated identities. It is expected that GET operations on this resource would be primarily for debugging purposes; thus, the representation should be human-readable. When human-readable, a returned MIME-type of *text/plain* is recommended. The representation should include a listing of the URLs for the delegated identities currently in scope on the server.

The representation of WR2, the delegated identity, must be the identity's distinguished name (DN) formatted as a string according to RFC 2253 [9] and returned with a MIME-type of *text/plain*.

> **Note:**
> For concerns of security and privacy, it is recommended that the representation of WR1, list of delegated identities *not* contain any information that can be used to ascertain the true identities of the users they represent. In particular, the WR1 representation should *not* list WR2 URLs if those URLs can be easily converted back into distinguished names. A unique but randomized name used in WR1 would be best at protecting privacy. If a disconnection between the WR2 URLs and the users' DNs cannot be made, the representation of WR1 could simply give the number of active identities in scope.

WR3, the certificate signing request (CSR), must be represented as a PKCS#10 [4] CSR with PEM encoding. ("PEM encoding" means that the text of the credential is written out as a byte stream according to the Distinguished Encoding Rules [7] of ASN.1 [8] and that stream is re-encoded in base 64.)

WR4, the proxy certificate, must be represented as an X.509v3 [5, 6] certificate with PEM encoding. More specifically, the certificate must be a proxy certificate following the rules of RFC 3820 [3]. The *proxyPolicy* field of the certificate must be set to the special value *id-ppl-inheritAll*, as defined in section 3.8.2 of RFC 3820.

> **Note:**
> The above *proxyPolicy* constraint means that the proxy identity represented by the certificate inherits all access rights of the identity from which it is derived; this kind of proxy certificate is colloquially called an "impersonation proxy".

> **Example:**
> Suppose, for example, that the client holds a certificate chain containing certificates with these subjects:
>
>     C=UK, O=AstroGrid, OU=Cambridge, CN=Guy Rixon
>     C=UK, O=AstroGrid, OU=Cambridge, CN=Guy Rixon, CN=12345678
>
> where the former is an EEC and the latter is a proxy certificate signed with the private key matching the EEC. The subject of the EEC is the delegated identity itself; this is the string returned as the representation of the identity resource. The certificate web resource will then be a further proxy certificate for, e.g.
>
>     C=UK, O=AstroGrid, OU=Cambridge, CN=Guy Rixon, CN=12345678, CN=9876543
>
> signed by the key matching the client's original proxy.

## 2.3 Operations on the resources

In this section, we detail the operations necessary for creating and using a proxy identity. In particular, we specify the required behavior of each of the operations. Figure 2 summarizes the complete sequence for the entire delegation process.

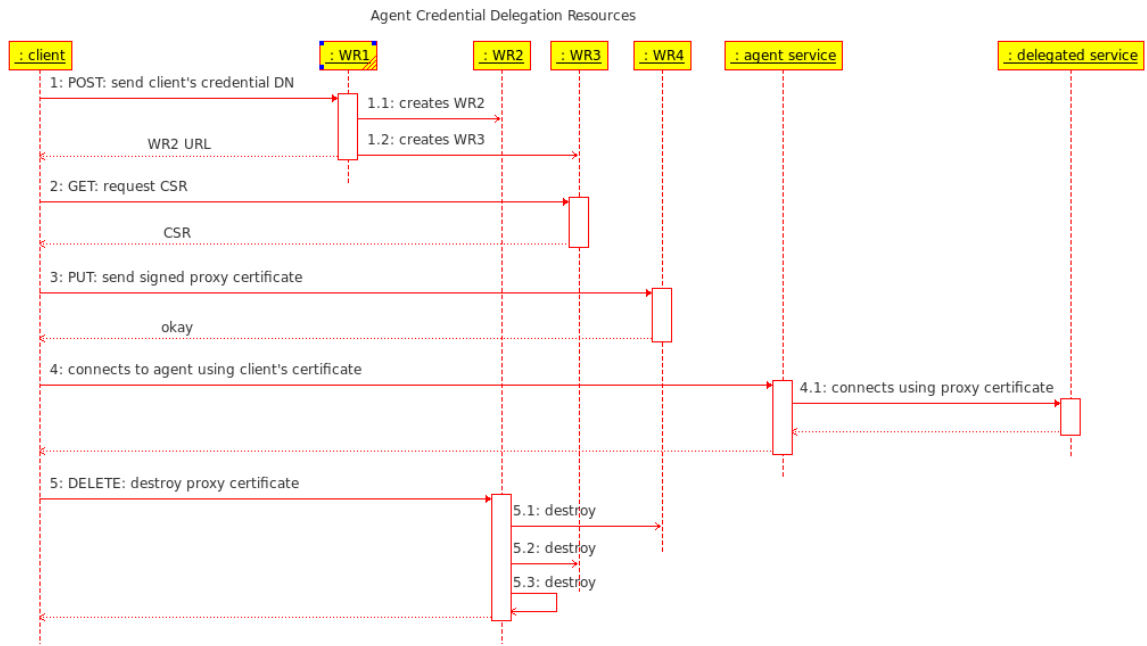

**Figure 2:** The sequence of operations for creating and using a proxy identity.

### 2.3.1 Establishing a delegated identity

Identity delegation is accomplished with the service through operations on the agent's web resources. The detailed specification of these operations is spelled out in section 2.3.3. In summary, these operations are, in order:

1. The client sends a POST request to WR1, sending the DN of the identity to be delegated. The agent creates WR2 and WR3 along with a public-private key pair, and responds with the newly created WR2 URL.

2. The client sends a GET request to the WR3 URL; the response is a Certificate Signing Request (CSR).

3. The client creates a signed proxy certificate from the CSR and then sends the certificate via a PUT request on the WR4 URL (thereby creating that resource, making its URL available for a GET request). The server stores the certificate for later use by agent.

(These three steps correspond to operations 1, 2, and 3 in Figure 2.)

At any time after the PUT operation, the client may request that the certificate be deleted (section 2.3.2; see also operation 5 in Figure 2). In the absence of such a request, it is left to the service implementer to decide for how long to keep the credentials stored.

> **Note:**
> Typical choices for how long to keep credentials include:
> - until the next restart of the service;
> - until the credentials expire and become invalid (as noted in the certificate);
> - forever.

Delegating new credentials for an existing identity (represented by an existing WR2 URL) must in effect replace the credentials.

### 2.3.2 Deleting a delegated identity and credentials

A delegated identity exists if there is a corresponding WR2 URL that is available responds to a respond to a GET request. A client can delete any delegated identity it has created by sending a DELETE request to the identity's WR2 URL. In response, the server will delete any stored proxy credentials for the identity and cause the WR2 URL, along with the URLs for its child resources (WR3 and WR4) to be unavailable.

The client may make such a DELETE request before issuing the PUT request to the associated WR4 URL. This in effect cancels the delegation process initiated with the first POST to WR1. Regardless, the WR2, WR3, and WR4 resources are deleted and their URLs become unavailable.

### 2.3.3 Operation specifications

All four types of web resources must respond to HTTP GET with the representations described above in section 2.2. Except for the requests described in the following sections, the agent must reject all other HTTP POST, PUT, and DELETE requests on any of the resources by returning the HTTP status, 403 "Forbidden". If any legal request should fail due to a problem on the server, it should return an appropriate HTTP status indicating a server error (i.e. a status of 500 or greater).

All the delegation web resources must be accessed via HTTPS; thus, their URLs must start with "https://". Authentication must use the *TLS-with-client-certificate method* [11]. The client certificate presented when the client-server connection is made identifies the base identity from which a proxy identity is to be created.

> **Note:**
> HTTPS also authenticates the server to the client. This requires that the client have access to the CA certificate that was used to sign the server's credentials in order for the connection to be successfully established. This document does not specify which CAs should be considered trustworthy by the client.

Successful operations on the WR2, WR3, and WR4 URLs must be *authorized* to ensure that the identity of the client is the same as the identity responsible for creating WR2. For the operation to be considered authorized, the delegated identity's DN that serves as the representation for WR2 (see section 2.2) must match the DN of the EEC component in the authenticating certificate chain presented by the client. If the DNs do not match, the agent must reject the operation request by returning an HTTP status of 403 "Forbidden".

### 2.3.3.1 WR1: the list of delegated identities

An authenticated HTTP GET request to the WR1 URL shall simply return the representation of WR1 as described in section 2.2 as the body of the response.

When the agent receives an authenticated HTTP POST to the WR1 URL, it shall create a delegated identity, represented by a new WR1 and based on the identity passed in the client's authenticating credentials. In particular, the agent must extract the subject DN from the EEC component of the authenticating certificate chain sent to the URL; this DN will serve as the representation of the delegated identity.

> **Note:**
> It is important that the agent use the EEC's DN, without extra proxy CN components, as the representation of the delegated identity. Since a client that authenticates with a proxy certificate will not typically use the same proxy certificate on every visit to the service (and thus the proxy DNs will not always have the same CN values), using the EEC DN allows the service to recognize a client that has visited the service previously.

On creating the delegated identity on the server, the agent shall create and store an RSA key pair. The agent shall also create the child web resource WR3 according to the specification in section 2.2. The agent shall return HTTP status 201 "Created" and shall include in the response the HTTP header named *Location* whose value is the absolute URL of WR2 representing the created identity.

### 2.3.3.2  WR2: the delegated identity

HTTP GET and DELETE operations must be supported on the WR2 URL. Successful operations on this URL must be authorized as described above in section 2.3.3.

An authorized HTTP GET request to the WR2 shall return the DN associated with the delegated identity as described in section 2.2.

An authorized HTTP DELETE request to the WR2 URL represents a request to delete the delegated identity from the server along with any associated credentials.  In particular, the agent shall delete the WR2 (by making its URL unavailable) along with all its child resources and the associated private key.

If the client makes a GET or DELETE request on the WR2 URL before a successful POST request has been made on WR1 or after a successful DELETE request on WR2 or after the agent has otherwise deleted the delegated identity, the agent shall return an HTTP status of 404 "Not Found" (since the WR2 does not exist).

### 2.3.3.3  WR3:  the certificate signing request

The WR3 URL must only support GET requests; a successful GET on this URL must be authorized as described above in section 2.3.3. With successful authorization, the agent shall return a Certificate Signing Request (CSR) to be converted to a properly signed proxy certificate by the client.  The certificate in the CSR must employ the public key that was created when the parent WR2 was created.

If the client makes a GET request on the WR3 URL before a successful POST request has been made on WR1 or after a successful DELETE request on WR2 or after the agent has otherwise deleted the delegated identity, the agent shall return an HTTP status of 404 "Not Found" (since the WR3 does not exist).

### 2.3.3.4  WR4:  the proxy certificate

HTTP GET and PUT operations must be supported on the WR4.  Successful operations on this URL must be authorized as described above in section 2.3.3.

An authorized HTTP PUT to the URL for WR4, the certificate resource of an identity, shall upload a proxy certificate that the agent may use on the client's behalf to access other secured services. The uploaded certificate shall be sent as the body of the request message.  This request creates the WR4 as a resource, making its URL available for GET requests. The agent shall store this certificate for its later use, and then respond to the client with an HTTP status of 201 "Created".

An authorized HTTP GET request to the WR4 URL shall return the uploaded certificate as the body of the response message using the representation specified in section 2.3.2. If a GET request to WR4 comes before a PUT request, the agent should return an HTTP status of 404 "Not Found" (since the resource does not technically exist until after a successful PUT).

## 2.4 Using the delegated credentials

The service which receives delegated credentials must authenticate the sender of any requests that would use those credentials. The sender must have the use of the identity from which the proxy credentials are derived or else the request shall be denied.

Delegated credentials created via the service interface described in section 2.3 must only be used within service interface implementations that are logically connected with the delegation service. For services that are registered in an IVOA-compliant registry [12], section 3 spells out how a VOResource-formatted description of the secure service indicates the application interfaces that require delegation via proxy credentials created with an associated delegation service.

Delegation credentials are valid only until the *use-by* time written into the proxy certificate; an attempt to use a proxy that has expired must result in an error.

The agent shall only make use of delegated proxy credentials in response to authenticated requests that are compliant with the IVOA standard for Authentication Mechanisms [1]. When a client makes such an request to a secure service requiring delegated credentials, the agent must choose the saved proxy credential corresponding to the delegated identity that matches the client's authenticated identity. In particular, the DN associated with the delegated identity must match the DN of the EEC component of the certificate chain used to authenticate the client. (This is the same technique described in section 2.3.3. used to determine authorization.) The agent must not use a proxy credential that does not match in this way. It is up to the secure service to determine how to otherwise react when matching delegated credentials cannot be found: it may fail with an error or continue operating without accessing any secondary secure services.

Section 3.1 indicates how to describe a delegation service in an IVOA resource registry and indicate which service interfaces that proxies created via the delegation service may be used. The implementation must restrict use of proxies to only those interfaces so indicated in the description. These interfaces must prevent access to the private keys by unauthorized agents, clients, or users.

> **Note:**
> Any process or user that has one of the private keys can, in principle, steal the corresponding delegated identity; thus, some care should be taken in storing them securely. In a Java implementation, the delegation resources and stored credentials might be managed by one servlet and the science service that uses them by another. These servlets would occur in the same virtual machine and can pass credentials as objects in memory, never committing them to disc. This is generally quite secure enough.
>
> If the science code cannot receive the credentials from memory, then precautions against unwanted copying are needed. Possible approaches include:
> - A shared file that is an encrypted key-store
> - Transfer via a MyProxy service, protected by TLS.

## 3 Registration of the service

A client can recognize a service that requires proxy credentials for delegation via its resource description in a resource registry [13]. When such a description is held in an IVOA-compliant registry (i.e. one that is compliant with the IVOA Recommendation for Registry Interfaces [13]), then its resource description will be in the form of a legal VOResource document [12] with the following requirements:

- The root resource element has an `xsi:type` attribute set to "*vr:Service*" or one of its legal extensions.

- The root resource element contains a `<capability>` child element with a `standardID` attribute set to "*ivo://ivoa.net/std/Delegation*" which describes the delegation interface. No `capability` extension (identified via an `xsi:type` attribute) is required.

- The delegation capability describing the delegation interface must contain an `<interface>` element with the `role` attribute set to "std"; that interface element must have an `<accessURL>` element whose value is the WR1 URL and which has a `use` attribute set to "*full*".

When the delegation interface is described this way, then the service must use proxy credentials created via that interface only within other the interface implementations that are described within that same VOResource Service description and which indicate that delegation is required. An interface indicates that delegation is required when the `<interface>` element includes a **`<securityMethod>`** child element having a **`standardID`** attribute set to "*ivo://ivoa.net/std/Delegation*".

> **Note:** a service resource indicates that a data access service requires delegated credentials by including a **`<securityMethod>`** element in the service's interface element:
>
> ```
> <capability>
>   <interface xsi:type="vs:ParamHTTP">
>     <accessURL use="base">http://a.b.c/d/getData</accessURL>
>     <securityMethod standardID="ivo://ivoa.net/std/Delegation"/>
>     <queryType>GET</queryType>
>     <resultType>text/xml</resultType>
>   </interface>
> </capabilty>
> ```
>
> The delegation service to use is described in the same resource record with its own capability description:
>
> ```
> <capability standardID="ivo://ivoa.net/std/Delegation">
>   <interface xsi:type="vs:ParamHTTP">
>     <accessURL use="full">http://a.b.c/d/delegations</accessURL>
>     <queryType>POST</queryType>
>     <resultType>text/plain</resultType>
>   </interface>
> </capabilty>
> ```

# Appendix A.  Change History

Since PR v1.0, 20090918:
- Added abbreviation glossary (App. B)
- Added figure
- Added page numbers

Since PR v1.01, 20080923:
- Re-arranged information for improved readability
- mandate *TLS-with-client-certificate* authentication only; drop mention of DN parameter.
- Clarified Error status responses
- An example of the various distinguished names in a delegation was added as commentary.

# Appendix B.  Glossary of Abbreviations

The following abbreviations are used in this document.  The page numbers indicate where the abbreviation is introduced.  Not included here are the acronyms defined in the preface section "Conformance related definitions.".

C – Country, a component of a distinguished name (DN) that identifies the country or origin for the thing being identified (p. 4).

CN – common name, a component of a distinguished name (DN) that identifies the particular person, server, or other identity within some organization.  The CN, when describing a person, will usually have the person's full name.  Additional CN components may further qualify the identity for a particular context (p. 4).

CSR – certificate signing request, a request to a party to create a certificate for a given identity and associated attributes and sign it with the private key held by the party (p. 5)

DN – distinguished name, the globally unique name for the identity described in an X.509 certificate (p. 4).

EEC – end-entity certificate, the X.509 certificate issued to a user intended to represent the user over the long-term.  In order to use it directly, the user must have the private key associated with the certificate.  An EEC contrasts against a proxy certificate which is short-lived (i.e. has a near-term validity expiration) and created from an EEC in that the proxy is signed with the EEC's private key (p. 4).

HTTPS – the HTTP protocol sent over the TLS protocol.  In order to authenticate a web service query using X.509 certificates, HTTPS must be used (p. 5).

MIME – Multipurpose Internet Mail Extensions, a standard for identifying used the type of data and (often) its format for the data being delivered in a message.  A "MIME-type" is a specific identifier used by HTTP to identify the specific type of data being passed (p. 7).

O – Organization, a component of a distinguished name (DN) that identifies the institutional organization that the thing being identified is part of (p. 4).

OU – Organizational unit, a component of a distinguished name (DN) that identifies the sub-group the thing being identified is a part of (p. 4).

PEM – Privacy Enhance Mail, a standard mechanism for encoding a data stream. X.509 certificates and certificate signing requests (CSR) are usually PEM encoded (p. 8).

PKCS – public key cryptography standards, a set of standards for securely packaging and encrypting information to be passed over the network. PKCS#10 refers to a specific standard [4] often used to encode a certificate signing request (CSR; p. 8).

REST – Representational state transfer, an approach to implementing and accessing web services based on four of the standard operations that are part of the HTTP protocol. In the VO, "RESTful" services are often contrasted with SOAP-based services (p. 5)

SOAP – Simple Object Access Protocol, a protocol for enabling remote procedure calls (execution of operations in software running on another machine). The messages are formatted as XML and can be sent over a number of underlying protocols, HTTP being the most common one (p. 4).

TLS – Transport Layer Security, a protocol for establishing a secure, authenticated connection between a client and a server. Based on the Secure Socket Layer (SSL), it authenticates both the client's and the server's identities using X.509 certificates presented by both sides (p. 3; see also [1], [2], and [3] for additional references and information).

URL – Universal Resource Locator, the identifier that identifies a specific web resource. It is intended that a URL can be resolved into a stream of data that represents the resource (p. 5).

WR1-4 – an identifier used (solely) in this report to indicate one of the four web resources that must be implemented to be compliant with this service standard (p. 6).

X.509 – a standard file/wire format for identity certificates based on public-private key encryption (p. 3)